\documentclass[twocolumn]{aastex631}

\usepackage{xcolor}
\usepackage{xspace}
\usepackage{amsmath}
\usepackage{CJKutf8}

\mathchardef\mhyphen="2D

\begin{document}
\begin{CJK*}{UTF8}{gbsn}
\title{Secular Dynamics of Compact Three-Planet Systems}
\author{Qing Yang (杨晴)}
\affiliation{Department of Physics, Harvey Mudd College, Claremont, CA 91711, USA}
\author{Daniel Tamayo}
\affiliation{Department of Physics, Harvey Mudd College, Claremont, CA 91711, USA}

\begin{abstract}
The secular Laplace-Lagrange orbital solution, decomposing eccentricities into a set of uniformly precessing eigenmodes is a classical result that is typically solved numerically.
However, in the limit where orbits are closely spaced, several simplifications make it possible to make analytical progress.
We derive simple expressions for the eccentricity eigenmodes in a co-planar 3-planet system where the middle planet is massless, and show that these approximate the true eigenmodes of more general systems with 3 massive planets in various limits. 
These results provide intuition for the secular dynamics of real systems, and have applications for understanding the stability boundary for compact multi-planet systems.

~\\ 
\end{abstract}

\section{Introduction}
\end{CJK*}

At low orbital eccentricities and inclinations, and away from mean motion resonances (MMRs), the evolution of eccentricities and inclinations in multiplanet systems behave approximately like a set of coupled harmonic oscillators. 
In an N-planet system, the seemingly complicated evolution of each planet's orbital eccentricity can thus be decomposed into $N$ eccentricity eigenmodes that precess at constant eigenfrequencies, with an independent set of $N$ inclination eigenmodes determining the inclination evolution \citep{Murray99}.
These secular modes have provided a valuable, approximately constant set of variables with which to explore our Solar System's more subtle evolution on longer timescales \citep[e.g.,][]{Laskar90, Mogavero21, Hoang21}, as well as a valuable lens through which to explore the range of possible dynamical behaviors in exoplanet systems where parameters are less constrained \cite[e.g.,][]{vanLaerhoven12, Zhang13}.

In this work, we explore these secular dynamics in the compact limit where orbits are tightly spaced. 
In addition to providing analytical insight for more widely spaced systems, this could have applications to planetary rings, debris disks and other such systems at the present day.

The compact limit is also important for understanding the onset of instabilities in young planetary systems, which we expect to form with closer orbital separations and undergo a giant impact phase \citep[e.g.,][]{Hansen13, Tremaine15, Dawson16} that sets the final masses and orbital architectures we observe today \citep[e.g,][]{Volk15, Pu15}.
\cite{Wisdom80} showed that the minimum stable orbital spacing for a pair of planets on initially circular orbits is set by the overlap of mean motion resonances (MMRs), a result that was extended to eccentric orbits by \cite{Hadden18}.
However, compact systems with more than two planets require significantly wider orbital spacings for stability and exhibit a much larger dynamic range of instability timescales \citep{Chambers96, Smith09, Obertas17, Lissauer21}.

\cite{Tamayo16} and \cite{Tamayo20} posited that even in compact systems with higher multiplicities, adjacent trios of planets provide the fundamental building block with which to understand stability, so this is the case we focus on in this work.
The two new effects are that, first, MMRs between two different pairs of planets can interact to produce chaotic regions in phase space through the overlap of 3-body resonances \citep{Quillen11, Petit20, Rath22}.
Second, \cite{Tamayo21} showed that the long-term secular oscillations in the eccentricities discussed above cause the widths of MMRs to adiabiatically breathe in and out, modulating the boundary at which MMRs overlap and drive widespread chaos.
This should in principle happen for two-planet systems too, but the particular combination of eccentricities that sets the MMR widths approximately coincides with one of the Laplace-Lagrange modes that is conserved, so this effect only appears for 3+ planet systems (see Sec.\:\ref{sec:2planets}).

This motivates better understanding the secular dynamics of 3-planet systems.
On one level, this problem was solved centuries ago.
However, the matrix diagonalization is typically performed numerically \citep{vanLaerhoven12}, because the general solution for the eigenvectors and eigenfrequencies in terms of the masses and orbital separations would span an entire page, and is thus of little use for applications.
To make analytical progress, \cite{Tamayo21} studied the dynamics in the limit where two closely spaced planets are perturbed by a distant third body.
However, this approximation is of limited use given that typical planetary systems have a tendency toward uniform spacings \citep{Weiss18}.

In this paper we seek more general results, driven by the expectation that the expressions should nevertheless simplify significantly in the compact limit. 
At close separations, where the eccentricities are necessarily small to avoid orbit-crossing, one can linearize the planets' unperturbed two-body paths following the ``guiding-center approximation" \citep{Murray99}.
This implies that only a particular linear combination of the eccentricity vectors determines the relative motion \citep{Namouni96}, leading to an additional conserved quantity \citep{Goldreich81, Henon86}.
This insight has provided significant simplification and intuition for MMR dynamics \citep{Hadden19}, and we show it similarly elucidates the secular problem.

We begin in Sec.\:\ref{sec:secular dynamics} by introducing our Hamiltonian approach and applying it to the two-planet case following \cite{Tamayo21}. 
We then use this intuition to guide the development of the 3-planet case in Sec.\:\ref{sec:3planets}.
We derive a solution in the compact limit where the middle planet is massless in Sec.\:\ref{sec:test particle}, and then explore how this result generalizes to wider separations in Sec.\:\ref{sec:generalization} and to massive middle planets in Sec.\:\ref{sec:generalizationmassive}.
We provide a sample application and comparison to N-body in Sec.\:\ref{sec:application} and conclude in Sec.\:\ref{sec:conc}.

\section{Secular Dynamics} \label{sec:secular dynamics}
At low inclinations typical of transiting planets, the stability of compact systems is not particularly sensitive to the inclination degrees of freedom \citep{Tamayo20, Tamayo21}. 
We therefore model the system as a co-planar.
Because the degrees of freedom corresponding to the semimajor axes are conserved in the secular problem \citep{Murray99}, there are $N$ degrees of freedom for the eccentricity of each planet to consider.

\subsection{Hamiltonian Laplace-Lagrange Formalism} \label{sec:formalism}
We adopt canonical Poincar{\'e} variables, with actions and conjugate angles given by
\begin{eqnarray}
\Gamma_i = m_i \sqrt{GM_\star a_{i}} \Bigg(1 - \sqrt{1-e_i^2}\Bigg) &\leftrightarrow& \gamma_i = -\varpi_i \label{canpoincare}
\end{eqnarray}
where $G$ is the gravitational constant, $M_\star$ the stellar mass, the $m_i$ are the planetary masses, and the $a_i$, $e_i$ and $\varpi_i$ are the orbital semimajor axes, eccentricities and logitudes of pericenter, respectively. To make our expressions easier to manipulate, we adopt the complex variables 
\begin{equation}
\boldsymbol{G_i} \equiv \sqrt{2\Gamma_i} e^{i\gamma_i}. \label{eq:G},
\end{equation}
and note that at low eccentricities,
\begin{equation}
\boldsymbol{G_i} \approx (GM_{\star}a_i)^{1/4}\sqrt{m_i}\,\boldsymbol{e_i^*},
\end{equation}
approximately proportional to the complex eccentricity
\begin{equation}
\mathbf{e}_i = e_i e^{i\varpi_i}, \label{eq:ei}
\end{equation}
which can be thought of as a vector of magnitude $e_i$ that points in the direction of pericenter $\varpi_i$ (in the complex plane).

Introducing a column vector of all the actions $\boldsymbol{G} \equiv (\boldsymbol{G_1}, ..., \boldsymbol{G_N})^T$, the Laplace-Lagrange Hamiltonian can be written as a compact matrix product,
\begin{equation}
\mathcal{H} = - \frac{1}{2}\boldsymbol{G}^T
 \cdot{\mathcal{M}}\cdot
\boldsymbol{G}^*,
\end{equation}
where $\mathcal{M}$ is an $N\times N$ matrix with real entries involving the planetary masses and Laplace coefficients (we derive the 3-planet case in Appendix \ref{app:laplag}, see Eq.\:\ref{eq:M}).

The $N$ equations of motion are then given by
\begin{equation}
    \frac{d}{dt}\boldsymbol{G} = -i\mathcal{M} \cdot \boldsymbol{G}.
\end{equation}

When expressed in terms of canonical variables (rather than orbital elements, e.g., \citealt{Murray99}), the Laplace-Lagrange matrix $\mathcal{M}$ becomes symmetric, making it clear through the spectral theorem that one can find a rotated basis in which the matrix is diagonal with real eigenvalues.

The diagonalizing rotation matrix $\mathcal{R}$ yields variables
\begin{equation}
    \boldsymbol{S} = \mathcal{R}\cdot\boldsymbol{G},\label{eq:rot_var}
\end{equation}
which are the Laplace-Lagrange modes, with magnitudes set by initial conditions that remain constant and rotate in phase at the corresponding eigenfrequency along the diagonal entries of $\mathcal{D} = \mathcal{R}\cdot\mathcal{M}\cdot\mathcal{R}^T$. 

This provides a valuable geometrical picture where the evolution of each $\boldsymbol{G_i}$ is given as a vector sum of contributions from these uniformly rotating modes (see Sec.\:\ref{sec:2planets} and Fig.\:\ref{fig:2planets}).

We will repeatedly exploit the fact that for compact systems, the ``center-of-mass eccentricity" $\boldsymbol{e_{com}} = \tilde{m_1}\boldsymbol{e_1} + \tilde{m_2}\boldsymbol{e_2}$ is conserved \citep{Goldreich81}, where throughout the paper we define $\tilde{m}_i \equiv m_i/m_{tot}$ as the fraction of the total \textit{planetary} mass.
This implies that $\boldsymbol{S}_{com} \propto \boldsymbol{e_{com}}$ is always an eigenmode of a compact $N$-planet system with zero eigenfrequency (i.e., its magnitude and direction are conserved).
This arises from a somewhat analogously in the two-body problem where the center-of-mass degree of freedom is conserved and one can reduce to an equivalent one-body problem.

\subsection{Compact 2-Planet Systems} \label{sec:2planets}

\begin{figure*}
\centering
\resizebox{0.7\textwidth}{!}{\includegraphics[]{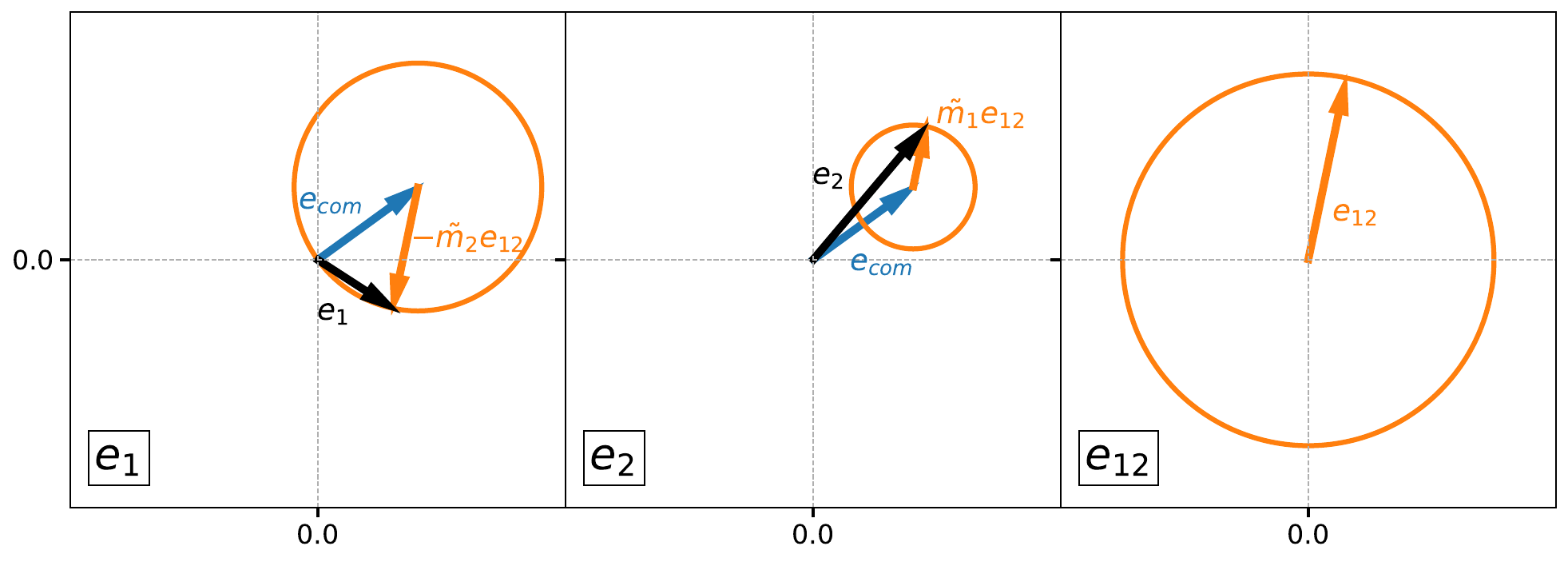}}
\caption{In a compact 2-planet system, the eccentricities $\boldsymbol{e_1}$ (left panel) and $\boldsymbol{e_2}$ (middle panel) can be decomposed as the vector sum of contributions from the two Laplace-Lagrange eigenmodes $\boldsymbol{e_{com}}$ (blue) and $\boldsymbol{e_{12}}$ (orange). The magnitude and direction of $\boldsymbol{e_{com}}$ are conserved so the blue arrows remain fixed, while the $\boldsymbol{e_{12}}$ mode rotates uniformly at fixed magnitude along the orange circles.
As this occurs, the magnitudes of $\boldsymbol{e_1}$ and $\boldsymbol{e_2}$ (black arrows) periodically grow and shrink, but because both share the same contribution from $\boldsymbol{e_{com}}$, their relative eccentricity $\boldsymbol{e_{12}}$ is conserved (right panel).
One can also see that the total $\boldsymbol{e_{12}}$ (right panel) gets split between the two eccentricities (left and middle panels) in proportion to the planetary masses. \label{fig:2planets}}
\end{figure*}

\cite{Tamayo21} show that in the compact limit for 2 planets, if we approximate the semimajor axes as equal,

\begin{eqnarray}
    \begin{pmatrix}\boldsymbol{S_{12}} \\ \boldsymbol{S_{com}} \end{pmatrix} &\approx& -
    \begin{pmatrix}
    \sqrt{\frac{m_2}{m_1+m_2}} & -\sqrt{\frac{m_1}{m_1+m_2}} \\
    \sqrt{\frac{m_1}{m_1+m_2}} & \sqrt{\frac{m_2}{m_1+m_2}}
    \end{pmatrix}
    \begin{pmatrix}
    \sqrt{m_1} & 0 \\ 0 & \sqrt{m_2}
    \end{pmatrix}
    \begin{pmatrix}\boldsymbol{e^*_1} \\ \boldsymbol{e^*_2}  \end{pmatrix} \nonumber \\ 
    &=& \begin{pmatrix} \sqrt{\frac{m_1m_2}{m_1+m_2}} \boldsymbol{e^*_{12}} \\ -\sqrt{m_1 + m_2} \boldsymbol{e^*_{com}} \end{pmatrix} \label{eq:2pmodes}
\end{eqnarray}
where $\boldsymbol{e_{12}} \equiv \boldsymbol{e_2} - \boldsymbol{e_1}$.
Inverting this equation to express the $\boldsymbol{G_i}$ as a sum of the constant modes becomes ill-defined in the test particle limit where one of the planet masses vanishes and $\boldsymbol{S_{12}} \rightarrow 0$.
The eccentricities nevertheless remain well behaved.
Since $\mathcal{R}$ is a rotation matrix (so $\mathcal{R}^{-1} = \mathcal{R}^T$), we have from Eq.\:\ref{eq:2pmodes}

\begin{eqnarray}
    \begin{pmatrix}\boldsymbol{e_1} \\ \boldsymbol{e_2} \end{pmatrix} &\approx&
    \begin{pmatrix}
    \frac{1}{\sqrt{m_1}} & 0 \\
    0 & \frac{1}{\sqrt{m_2}}
    \end{pmatrix}
    \mathcal{R}^{T}
    \begin{pmatrix}
    \sqrt{\frac{m_1m_2}{m_1+m_2}} & 0 \\
    0 & -\sqrt{m_1 + m_2}
    \end{pmatrix}
    \begin{pmatrix}\boldsymbol{e_{12}} \\ \boldsymbol{e_{com}} 
    \end{pmatrix} \nonumber \\
    &=& \begin{pmatrix} -\tilde{m}_2 \boldsymbol{e_{12}}(t) + \boldsymbol{e_{com}} \\ \tilde{m}_1 \boldsymbol{e_{12}}(t) +\boldsymbol{e_{com}} \end{pmatrix}. \label{eq:2pdecomp}
\end{eqnarray}

Equation \ref{eq:2pdecomp} provides the geometrical interpretation described at the end of Sec.\:\ref{sec:formalism}, and is shown in Fig.\:\ref{fig:2planets}.
Both $\boldsymbol{e_1}$ and $\boldsymbol{e_2}$ have equal contributions of $\boldsymbol{e_{com}}$, which is conserved and maintains constant magnitude and direction.
By contrast, the second mode\footnote{For ease of discussion, we will interchangeably refer to the mass-weighted canonical actions $\boldsymbol{S}$ and their corresponding unweighted eccentricities (e.g., $\boldsymbol{S_{12}}$ and $\boldsymbol{e_{12}}$) as ``modes".} $\boldsymbol{e_{12}} = \boldsymbol{e_{12,0}}e^{i\omega_{12} t}$ undergoes uniform rotation along the orange circle at a rate \citep{Tamayo21}
\begin{equation}
\omega_{12} \approx \frac{m_{tot}}{M_\star}\frac{1}{e_{c,12}^2}\frac{1}{P_2}, \label{eq:omega}
\end{equation}
where $P_2$ is the outer planet's orbital period, and $e_{c,12}$ is the value of $e_{12}$ at which the orbits would cross.
For tightly packed orbits
\begin{equation}
e_{c,ij} = 1 - \alpha_{ij} = 1 - a_i/a_j
\end{equation}
the secular timescale $T = 2\pi/\omega_{12}$ is typically quoted as $\sim (M_\star/m_{tot})P_2$ as appropriate for large crossing eccentricities (fractional separations), but we see that for compact systems, $T$ is significantly reduced by $e_{c,12}^2$.

If we subtract both rows of Eq.\:\ref{eq:2pdecomp}, the common $\boldsymbol{e_{com}}$ contribution cancels, and we self-consistently recover that $\boldsymbol{e_2}-\boldsymbol{e_1} = \boldsymbol{e_{12}}$.
Thus, the relative eccentricity $\boldsymbol{e_{12}}$ sweeps out a circle of constant radius, so the magnitude $e_{12}$ is conserved (right panel of Fig.\:\ref{fig:2planets}).

This is a somewhat redundant argument given that we already found in Eq.\:\ref{eq:2pmodes} that $\boldsymbol{e_{12}}$ was one of the modes (whose magnitudes are always conserved).
But the fact that the relative eccentricity $\boldsymbol{e_{12}}$ corresponds to one of the Laplace-Lagrange modes is specific to the compact 2-planet case, where it has the important implication that there is no secular modification to MMR widths \citep{Tamayo21}.

In general, for compact N-planet systems with $N>2$, it is no longer true that the relative eccentricities setting the MMR widths correspond to Laplace-Lagrange modes. 
However, we will show that the conservation of $\boldsymbol{e_{com}}$ implies that all the eccentricities have approximately equal contributions of the $\boldsymbol{e_{com}}$ mode, implying that in general, any relative eccentricities can only be made up of combinations of the remaining $N-1$ modes.

\section{Compact 3-Planet Systems} \label{sec:3planets}
\subsection{Test Particle Limit} \label{sec:test particle}
In the limit where the third planet in the middle is much less massive, the two massive planets still approximately undergo two-planet dynamics, so we know that two of the three eigenmodes are approximately those given above, i.e., $\boldsymbol{e_{com}}$ and $\boldsymbol{e_{13}}$ (note the changed index due to the additional middle planet).
These specify the bottom two rows of our rotation matrix $\mathcal{R}_1$.
The requirement that $\mathcal{R}_1$ be orthogonal and have a determinant of unity in order to be a rotation uniquely determines the third eigenmode along the top row,
\begin{equation}
\mathcal{R}_1=\begin{pmatrix}
\sqrt{\frac{\tilde{m}_1\tilde{m}_2}{\tilde{m}_1+\tilde{m}_3}} & -\sqrt{\tilde{m}_1+\tilde{m}_3} & \sqrt{\frac{\tilde{m}_2\tilde{m}_3}{\tilde{m}_1+\tilde{m}_3}}\\
-\sqrt{\frac{\tilde{m}_3}{\tilde{m}_1+\tilde{m}_3}} & 0 & \sqrt{\frac{\tilde{m}_1}{\tilde{m}_1+\tilde{m}_3}} \\
\sqrt{\tilde{m}_1} & \sqrt{\tilde{m}_2} & \sqrt{\tilde{m}_3}
\end{pmatrix}, \label{eq:R1}
\end{equation}
which yields three rotated variables
\begin{equation}
\begin{pmatrix}\boldsymbol{S'_1} \\ \boldsymbol{S'_2} \\ \boldsymbol{S'_3}\end{pmatrix} = 
\sqrt{m_{tot}}\begin{pmatrix}
\sqrt{\frac{\tilde{m}_2}{\tilde{m}_1+\tilde{m}_3}}\left(\tilde{m}_3\, \boldsymbol{e_{23}^*} - \tilde{m}_1\,\boldsymbol{e_{12}^*}\right)\\
\sqrt{\frac{\tilde{m}_1\tilde{m}_3}{\tilde{m}_1+\tilde{m}_3}}\,\boldsymbol{e_{13}^*}\\
\boldsymbol{e_{com}^*}
\end{pmatrix} \label{eq:S'}
\end{equation}

The new Hamiltonian is
\begin{equation}
\begin{gathered}
    \tilde{\mathcal{H}}= \begin{pmatrix}\boldsymbol{S'_1} & \boldsymbol{S'_2} & \boldsymbol{S'_3}\end{pmatrix} \cdot \mathcal{M'}\cdot \begin{pmatrix}\boldsymbol{{S'_1}^{*}} \\ \boldsymbol{{S'_2}^{*}} \\ \boldsymbol{{S'_3}^{*}}\end{pmatrix}, \\
    \mathcal{M'} = \mathcal{R}_1\cdot\mathcal{M}\cdot\mathcal{R}_1^{T}, \label{eq:new Hamiltonian}
\end{gathered}
\end{equation}
where $\mathcal{M'}$ is the rotated Laplace-Lagrange matrix.

In the compact limit where $\alpha$ is close to unity, the Laplace coefficients in Eq.\:\ref{eq:M} can be approximated as \citep{Tamayo21}
\begin{equation}
    b^{(m)}_{3/2}(\alpha) \approx \frac{2}{\pi(1-\alpha)^2},
\end{equation}
which is independent of $m$. We define a parameter $\delta$ to represent the fractional difference between the Laplace coefficients:
\begin{equation}
    \delta \equiv \frac{b^{(1)}_{3/2}(\alpha_{ij}) - b^{(2)}_{3/2}(\alpha_{ij})}{b^{(1)}_{3/2}(\alpha_{ij})}. \label{eq:delta}
\end{equation}

Substituting Eq.\:\ref{eq:delta} into Eq.\:\ref{eq:new Hamiltonian}, we can write the rotated Laplace-Lagrange matrix as
\begin{equation}
    \mathcal{M'} = \begin{pmatrix}\omega'_1 & k & 0\\
    k & \omega'_2 & 0\\
    0 & 0 & 0 \end{pmatrix} + \delta\omega'_1\mathcal{M}_d, \label{eq:M'}
\end{equation}
where we show in Appendix \ref{app:Mprime} the expressions for $\omega'_1$, $\omega'_2$ and $k$. The elements in $\mathcal{M}_d$ are $\mathcal{O}(1)$ or smaller, while $\delta \lesssim 0.1$ for $\alpha > 0.75$ and drops to zero as $\alpha$ approaches unity.
We therefore drop this additional correction.

The fact that we know that the eigenmodes must be approximately given by the rows of $\mathcal{R}_1$ (Eq.\:\ref{eq:R1}) implies that $k$ is much smaller than either of the $\omega$ along the diagonal. 
Nevertheless, we now show that it is important to perform one more rotation to remove these small off-diagonal terms.
This is easily achieved through a rotation
\begin{equation}
\begin{gathered}
    \mathcal{R}_2 = \begin{pmatrix}
    \cos{\psi} & -\sin{\psi} & 0 \\ \sin{\psi} & \cos{\psi} & 0 \\ 0 & 0 & 1
\end{pmatrix},
\end{gathered} \label{eq:R'}
\end{equation}
where
\begin{equation}
    \psi = \frac{1}{2}\tan^{-1}\left(\frac{2k}{\omega'_2-\omega'_1}\right).\label{eq:psi}
\end{equation}
The corresponding diagonal matrix $\mathcal{M''} = \mathcal{R}_2\cdot\mathcal{M'}\cdot\mathcal{R}_2^T$, with eigenmodes and eigenfrequencies given by
\begin{equation}
    \begin{pmatrix}\boldsymbol{S_1} \\ \boldsymbol{S_2} \\ \boldsymbol{S_3}\end{pmatrix} = 
    \begin{pmatrix}\cos\psi\boldsymbol{S'_1} - \sin\psi\boldsymbol{S'_2}\\ \sin\psi\boldsymbol{S'_1} + \cos\psi\boldsymbol{S'_2}\\
    \boldsymbol{S'_3}\end{pmatrix}, \label{eq:mode_general}
\end{equation}
\begin{equation}
    \begin{pmatrix}\omega_1 \\ \omega_2 \\ \omega_3\end{pmatrix} = 
    \begin{pmatrix}
        \omega'_1\cos^2{\psi} + \omega'_2\sin^2{\psi} - k\sin{2\psi}\\
        \omega'_1\sin^2{\psi} + \omega'_2\cos^2{\psi} + k\sin{2\psi}\\
        0
    \end{pmatrix}. \label{eq:freq_general}
\end{equation}

Equation \ref{eq:mode_general} helps clarify the issue. 
Even though $\psi \ll 1$ for a small middle planet, we have from Eq.\:\ref{eq:S'} that $S'_2$ is larger than $S'_1$ by a factor $\sim \sqrt{m_3/m_2} \gg 1$, so the two contributions to $\boldsymbol{S_1}$ above are in fact of the same magnitude, while $\boldsymbol{S_2} = \boldsymbol{S'_2}$ to excellent approximation.

\begin{figure*}
\centering
\resizebox{0.7\textwidth}{!}{\includegraphics[]{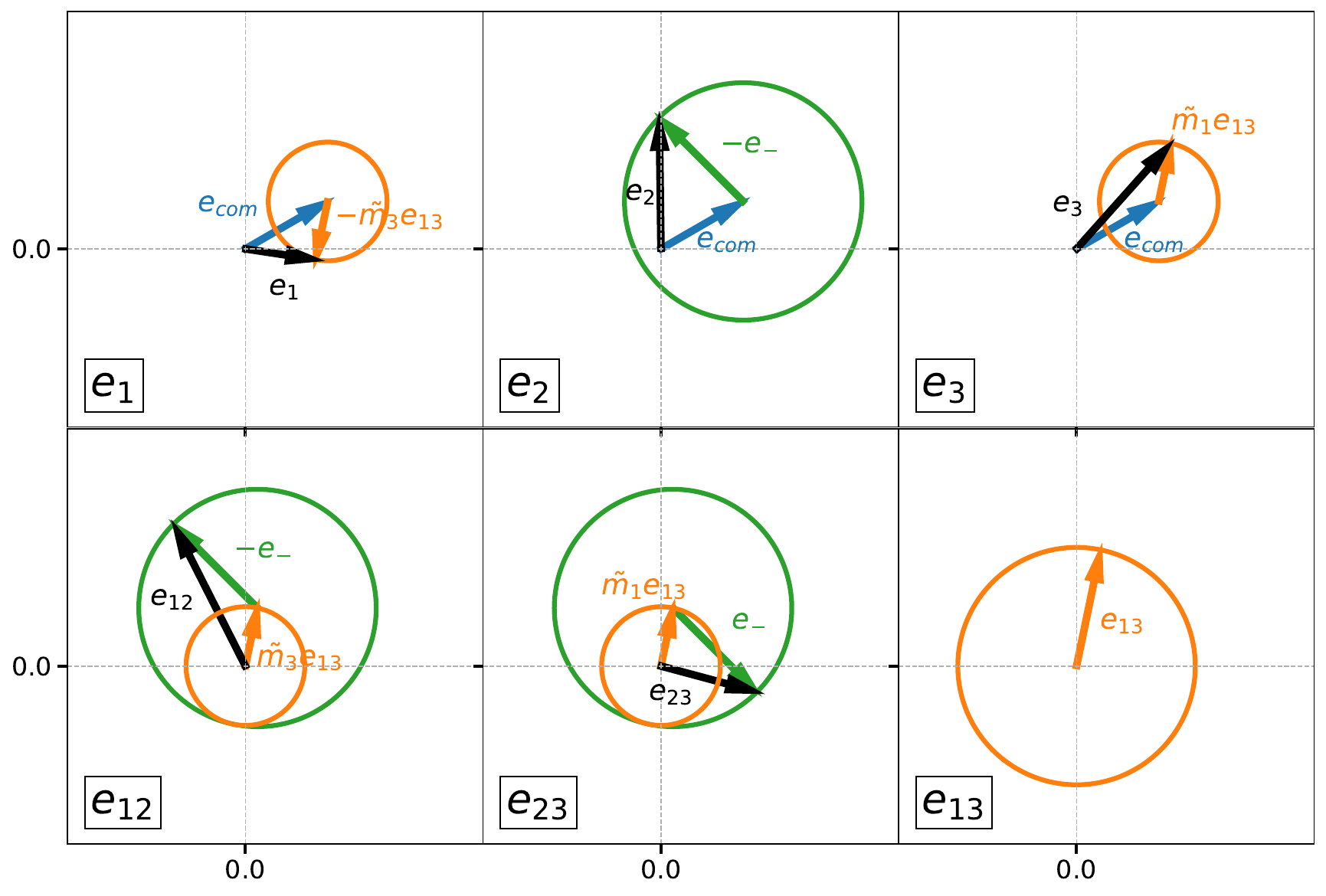}}
\caption{Analogous plot to Fig.\:\ref{fig:2planets}, for a compact 3-planet system where the middle planet is massless.
The two massive planets then follow 2-planet dynamics (compare the top left, top right and bottom right panels to Fig.\:\ref{fig:2planets}).
We see that the middle test particle's eccentricity is governed by the new $\boldsymbol{e_{-}}$ mode (green) with no contribution from $\boldsymbol{e_{13}}$ (orange).
Again, all eccentricities share the same contribution from $\boldsymbol{e_{com}}$ (blue), so the relative eccentricities in the bottom row of panels are independent of this eigenmode.
However, unlike in the 2-planet case, the relative eccentricities between adjacent planets $\boldsymbol{e_{12}}$ (bottom left panel) and $\boldsymbol{e_{23}}$ (bottom middle) are now in general governed by two eigenmodes and undergo oscillations, unlike the two-planet case. \label{fig:3planets}}
\end{figure*}

In this test particle limit, plugging Eqs.\:\ref{eq:w1}-\ref{eq:k} into Eq.\:\ref{eq:psi} for the angle $\psi$ yields
\begin{equation}\begin{aligned}
    \psi &\approx \frac{k}{\omega'_2-\omega'_1}\\ 
    &\approx \sqrt{\tilde{m}_1\tilde{m}_2\tilde{m}_3}
    \left(\tfrac{\tfrac{1}{e_{c,23}^{2}}-\tfrac{1}{e_{c,12}^{2}}}{\tfrac{\tilde{m}_1\tilde{m}_2-\tilde{m}_3}{e_{c,23}^{2}} + \tfrac{\tilde{m}_2\tilde{m}_3-\tilde{m}_1}{e_{c,12}^{2}} + \tfrac{(\tilde{m}_1+\tilde{m}_3)^2}{e_{c,13}^{2}}}\right). \label{eq:saa}
\end{aligned}\end{equation}

In the compact limit where all $e_c \ll 1$, we can additionally make the approximation that $e_{c,12} + e_{c,23} = e_{c,13}$. We define
\begin{equation}
    \eta \equiv \frac{e_{c,12}}{e_{c,13}} - \frac{e_{c,23}}{e_{c,13}},
\end{equation}
which acts as a spacing asymmetry parameter. For fixed inner and outer planet orbits, $\eta = -1$ corresponds to placing the middle planet's orbit all the way up against the inner planet's orbit, $\eta = +1$ up against the outer planet's orbit, and $\eta=0$ at equal spacing. This also implies
\begin{equation}
    \frac{e_{c,12}}{e_{c,13}} = \frac{1}{2}(1+\eta),\quad \frac{e_{c,23}}{e_{c,13}} = \frac{1}{2}(1-\eta).
\end{equation}
We also define a parameter that accounts for the asymmetry in the mass distribution of the inner and outer planets:
\begin{equation}
    \mu \equiv \frac{\tilde{m}_3 - \tilde{m}_1}{\tilde{m}_1 + \tilde{m}_3}.
\end{equation}
Then Eq.\:\ref{eq:saa} reduces to
\begin{equation}
    \psi = -\frac{8\sqrt{\tilde{m}_2(1-\mu^2)}\,\eta}{3+6\eta^2+8\mu\eta-\eta^4}
\end{equation}

When $\tilde{m}_2 \ll 1$, we have $\psi\ll 1$ and $|\boldsymbol{S'_1}|\ll 1$ in Eq.\:\ref{eq:mode_general}, so $\sin \psi \approx \psi$ and $\cos \psi \approx 1$. After significant algebra, the three eigenmodes can be expressed as
\begin{equation}
    \begin{pmatrix}\boldsymbol{S_{-}} \\ \boldsymbol{S_{+}} \\ \boldsymbol{S_{com}}\end{pmatrix} = \sqrt{m_{tot}}
    \begin{pmatrix}
    \sqrt{\tilde{m}_2}\,\boldsymbol{e_{-}^*}\\ 
    \sqrt{\tilde{m}_1\tilde{m}_3}\,\boldsymbol{e_{13}^*}\\
    \boldsymbol{e_{com}^*}\end{pmatrix}, \label{eq:mode_tp}
\end{equation}
where
\begin{equation}\begin{gathered}
\boldsymbol{e_{-}} \equiv \frac{\chi_{23}\,\boldsymbol{e_{23}} -\chi_{12}\,\boldsymbol{e_{12}}}{\chi_{23}+\chi_{12}},\\
\chi_{23} = (1+\eta)^3(3-\eta)\tilde{m}_3,\\
\chi_{12} = (1-\eta)^3(3+\eta)\tilde{m}_1.
\label{eq:eminus}
\end{gathered}\end{equation}
When the middle planet is much closer to the inner planet ($\eta \to -1$), the close pair forms a 2-planet ``subsystem", and $\boldsymbol{e_{-}} \to \boldsymbol{e_{12}}$.
Similarly for $\eta \to +1$, $\boldsymbol{e_{-}} \to \boldsymbol{e_{23}}$. 
For the evenly-spaced case where $\eta=0$, $\boldsymbol{e_{-}}$ reduces to $\tilde{m}_3\boldsymbol{e_{23}} - \tilde{m}_1\boldsymbol{e_{12}}$.

We can relate the eccentricity of each planet to the eigenmodes by $\boldsymbol{G} = (\mathcal{R}_2\mathcal{R}_1)^T \cdot \boldsymbol{S}$, which, following the procedure in Sec.\:\ref{sec:2planets} to convert to eccentricities, results in
\begin{equation}
    \begin{pmatrix}
        \boldsymbol{e_1}\\ \boldsymbol{e_2} \\ \boldsymbol{e_3}
    \end{pmatrix} = 
    \begin{pmatrix}
        \boldsymbol{e_{com}} - \tilde{m}_3\,\boldsymbol{e_{13}}\\
        \boldsymbol{e_{com}} - \boldsymbol{e_{-}}\\
        \boldsymbol{e_{com}} + \tilde{m}_1\,\boldsymbol{e_{13}}\\
    \end{pmatrix}, \label{eq:StoE}
\end{equation}
and
\begin{equation}
    \begin{pmatrix}
        \boldsymbol{e_{12}}\\ \boldsymbol{e_{23}}
    \end{pmatrix} = 
    \begin{pmatrix}
        \tilde{m}_3\,\boldsymbol{e_{13}} - \boldsymbol{e_{-}}\\
        \tilde{m}_1\,\boldsymbol{e_{13}} + \boldsymbol{e_{-}}\\
    \end{pmatrix}. \label{eq:StoErelative}
\end{equation}

We illustrate this test particle solution in Fig.\:\ref{fig:3planets}, analogous to Fig.\:\ref{fig:2planets}.
The top row of panels show the individual eccentricities as a vector sum of their constituent eigenmodes. 
All of them share the same contribution from $\boldsymbol{e_{com}}$, so this eigenmode disappears from the relative eccentricities along the bottom row of panels.
Unlike in the 2-planet case, however, the relative eccentricities between adjacent planets $\boldsymbol{e_{12}}$ and $\boldsymbol{e_{23}}$ are in general made up of the remaining two eigenmodes, causing the relative eccentricities to oscillate in time (see top panel of Fig.\:\ref{f:Nbody}, corresponding to the same system). 

We now explore how well these eigenmodes derived in the compact test-particle limit generalize to wider separations and massive middle planets.

\subsection{Generalization to Wider Spacings} \label{sec:generalization}
\begin{figure}
\includegraphics[width=1\columnwidth]{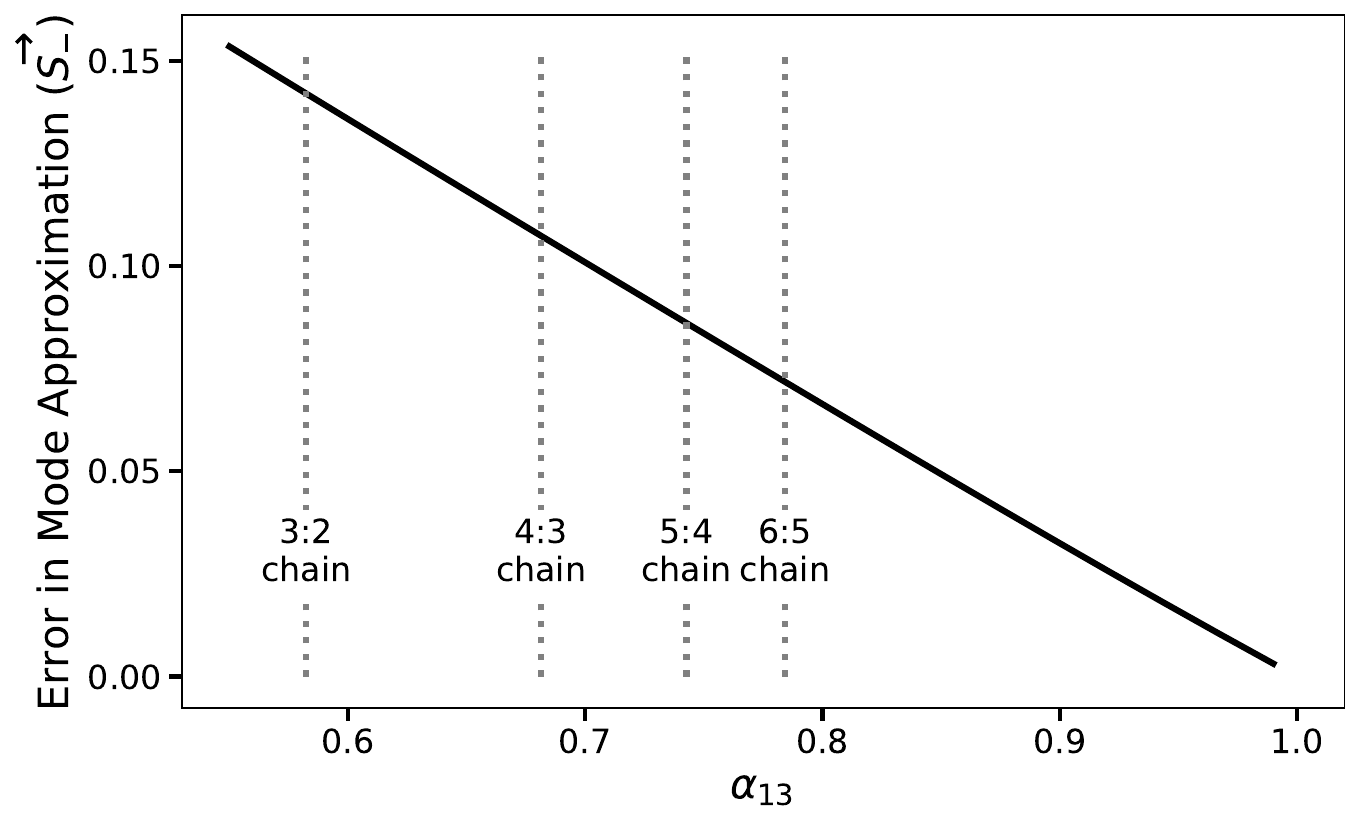}
\caption{Error (as defined in the text) for our compact approximation of the $\boldsymbol{e_-}$ eigenmode (Eq.\:\ref{eq:eminus}), considering an equally-spaced ($\eta=0$) 3-planet system with a massless middle planet.
On the $x$-axis we vary the ratio between the innermost and outermost semimajor axes, so that the compact approximation worsens moving left.
Vertical dotted lines correspond to separations where the period ratios between adjacent planets falls on different first-order MMRs.
The error is $\lesssim 15\%$ for period ratios between adjacent planets closer than the $3:2$.}
\label{f:varyAlpha13}
\end{figure}

We first explore the effect of wider spacings by varying $\alpha_{13} = a_1/a_3$ along the x axis, considering uniformly spaced planets $\eta = 0$ with a massless middle planet.
To quantify the error, we use the \texttt{celmech} package \citep{Hadden22} to numerically evaluate our Laplace-Lagrange matrix (Eq.\:\ref{eq:M}).
We then diagonalize it numerically to obtain the true $\boldsymbol{e_-}$, and normalize it to a unit vector.
We then calculate our approximated unit vector for $\boldsymbol{e_-}$ using Eq.\:\ref{eq:eminus}, and calculate the distance between our approximated unit vector and the true one.
We show the result in Fig.\:\ref{f:varyAlpha13}, overplotting some separations corresponding to MMRs between adjacent planets.
We see that for closer separations than the $3:2$ (between adjacent planets), the error is $\lesssim 15\%$.

\subsection{Generalization to Massive Middle Planets} \label{sec:generalizationmassive}
\begin{figure}[h]
\includegraphics[width=1\columnwidth]{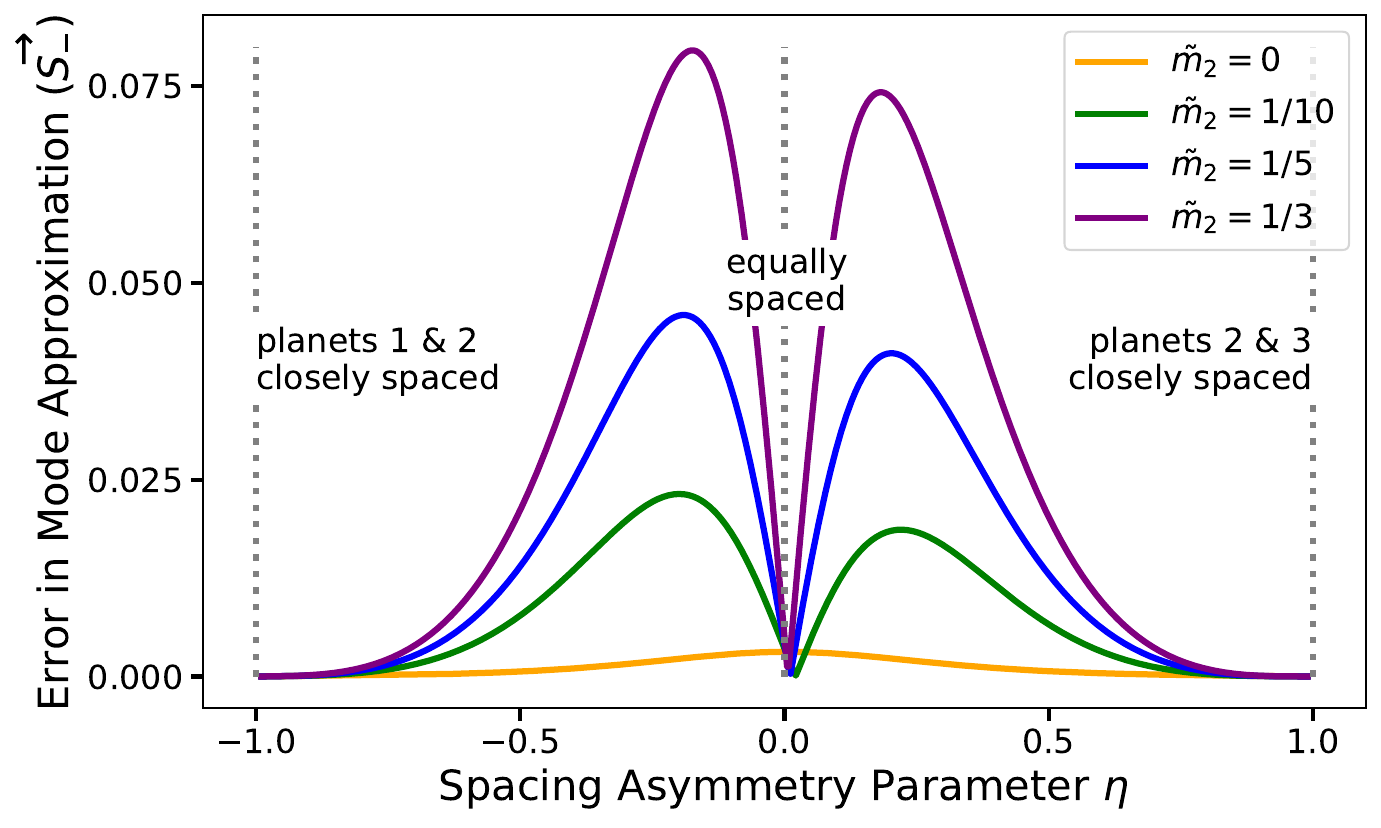}
\caption{Error for our test-particle approximation of the $\boldsymbol{e_-}$ eigenmode (Eq.\:\ref{eq:eminus}) as we increase the mass of the middle planet from the test-particle case (orange curve) to the equal-mass case (purple curve).
To isolate only errors in our test particle approximation, we set $\alpha_{13}=0.99$ so that the errors introduced by the compact approximation are negligible (Fig.\:\ref{f:varyAlpha13}). 
We plot the error as a function of the spacing asymmetry parameter $\eta$, so the left of the plot corresponds to placing the middle planet right next to its inner neighbor (and vice versa on the right of the plot), while $\eta=0$ corresponds to uniform spacing. \label{f:varyEta}}
\end{figure}

When the middle planet is massive, there are additional terms $\propto \tilde{m}_2$ in the rotational angle $\psi$ (Eq.\:\ref{eq:psi}) given by Eqs.\:\ref{eq:w1}-\ref{eq:k}.
If the rotation angle is still small, $\cos{\psi} \approx 1$ and $\sin{\psi} \approx \psi$ with
\begin{equation}
\psi\approx -\frac{8\sqrt{\tilde{m}_2(1-\mu^2)}\eta}{(3+6\eta^2+8\mu\eta-\eta^4) - \tilde{m}_2(3+6\eta^2-8\mu\eta-\eta^4)}.
\end{equation}
After signficant algebra, this yields a corrected mode $\boldsymbol{e_-^{(1)}}$
\begin{equation}
    \boldsymbol{e_-^{(1)}} = \boldsymbol{e_-} + \frac{4}{3}\tilde{m}_2\,\eta\,[(1+\mu)^2\,\boldsymbol{e_{23}} + (1-\mu)^2\,\boldsymbol{e_{12}}]. \label{eq:correction}
\end{equation}
As expected, the correction vanishes as $\tilde{m}_2 \to 0$ in the test particle limit.
Surprisingly, however, it also vanishes as $\eta \to 0$.
Uniformly spaced 3-planet systems thus share the same eigenmodes as the test particle case.
This is visible in Fig.\:\ref{f:varyEta}, where we plot the error in our test particle mode (Eq.\:\ref{eq:eminus}) as a function of the spacing asymmetry parameter $\eta$, with different color curves corresponding to different masses for the middle planet.
To isolate this effect, we consider $\alpha_{13} = 0.99$ where the errors from the compact approximation are negligible (Fig.\:\ref{f:varyAlpha13}).
In orange we plot the test particle case, which has small errors throughout. 
As the middle planets becomes progressively more massive (reaching the equal-mass case in purple), the errors grow in a V shape at small $\eta$, corresponding to the $\tilde{m}_2 \eta$ scaling of the correction in Eq.\:\ref{eq:correction}.
If we include the correction, we find that we indeed fix the errors for $|\eta| \lesssim 0.1$; however, at larger $\eta$, our small-angle approximation for $\psi$ becomes poor, and including this correction gives the wrong asymptotic behavior as $|\eta| \to 1$. 

\begin{figure}[h]
\includegraphics[width=1\columnwidth]{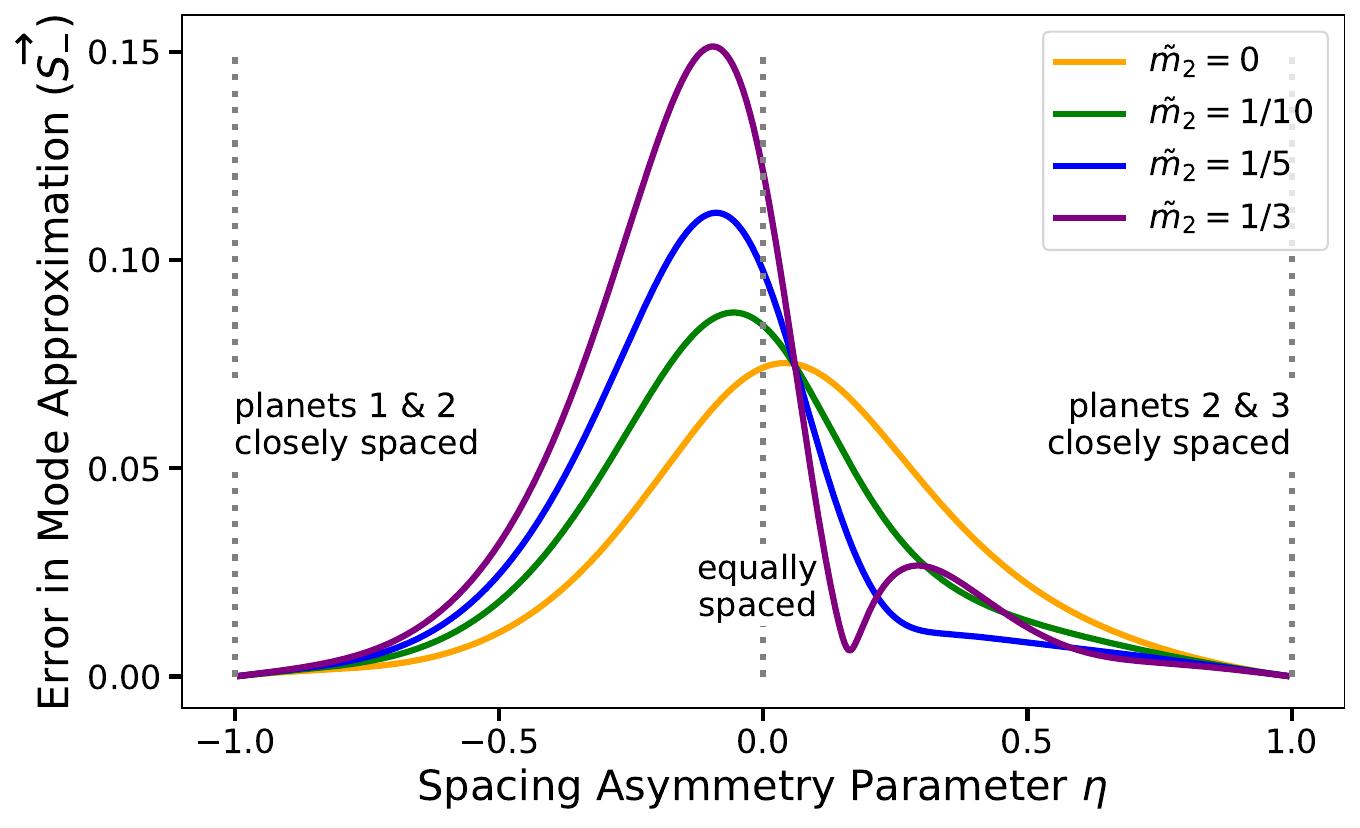}
\caption{We loosen the compact limit constraint by remaking Fig.\:\ref{f:varyEta} with a more widely spaced $\alpha_{13}=0.78$. The errors are now due both to wider spacings and a massive middle planet, leading to a more complicated pattern skewed towards negative $\eta$, and with non-zero error even in the test-particle case.}
\label{f:varyEta_noncompact}
\end{figure}

We conclude that the test particle result (Eq.\:\ref{eq:eminus}) provides not only a simple expression, but also one with the correct asymptotic behavior of approaching $\boldsymbol{e_{12}}$ (or $\boldsymbol{e_{23}}$) when the middle planet is very close to its inner (or outer) neighbor.

The behavior when there are errors both due to wider separations and a massive middle planet is more complicated, since the errors can add or partially cancel in a variety of ways. 
To illustrate this, we remake Fig.\:\ref{f:varyEta} for a wider total spacing $\alpha_{13} = 0.78$. 
We see that the error pattern in Fig.\:\ref{f:varyEta_noncompact} becomes skewed, and even the test-particle case has errors due to the wider separations.

\section{An Application} \label{sec:application}

\begin{figure}[h]
\includegraphics[width=1\columnwidth]{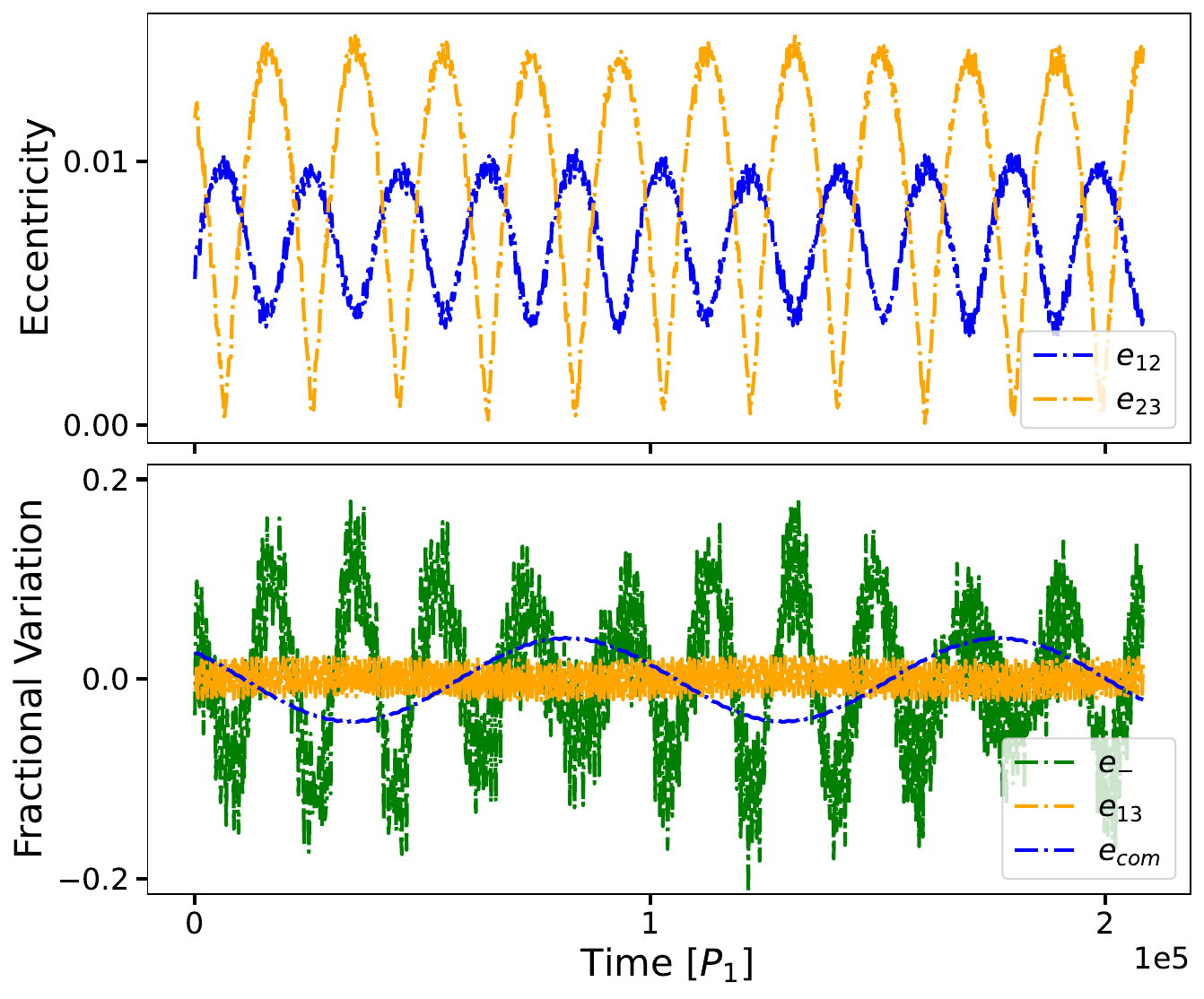}
\caption{We consider a pair of Earth-mass planets with period ratio 1.54 wide of the 3:2 MMR, with a test particle placed approximately in the middle ($\eta = -0.12$). The eccentricities and pericenters are drawn to match the mode amplitudes and evolution depicted in Fig.\:\ref{fig:3planets}). 
Top panel plots the relative eccentricities vs time.
In the bottom panel, we plot the relative variations of our approximated modes vs time, which would ideally remain at zero.
We see that the relative variations remain $\lesssim 20\%$.
\label{f:Nbody}}
\end{figure}

Finally, we compare these results directly against N-body integration.
We consider two Earth-mass planets with period ratio 1.54, wide of the 3:2. 
We then insert an additional test particle with $\eta = -0.12$, corresponding to approximately uniform spacing.
In the top panel of Fig.\:\ref{f:Nbody} we plot the variations of the magnitudes of $\boldsymbol{e_{12}}$ and $\boldsymbol{e_{23}}$ vs. time. 
We see that they are out of phase with one another, which is a generic result of the test particle limit due to the opposite signs of $\boldsymbol{e_-}$ in the relative eccentricities in Eq.\:\ref{eq:StoErelative}. This is reflected geometrically in the opposite directions of the green vectors in the two bottom left panels for the relative eccentricities in Fig.\:\ref{fig:3planets}, corresponding to the same planetary system.

In the bottom panel of Fig.\:\ref{f:Nbody} we plot the time evolution of our three approximated modes as a fractional deviation from their means, i.e., $[e(t) - \bar{e}]/\bar{e}$, where the bars denote mean values.
We see that the fractional variations in $\boldsymbol{e_-}$ are $\lesssim 20\%$, while the variations in the remaining modes are significantly smaller.

\section{Conclusion} \label{sec:conc}

We have explored how the secular evolution of multiplanet systems breaks down into simply understood Laplace-Lagrange eigenmodes in the limit where the orbits are closely spaced.
In this compact limit, the center-of-mass eccentricity vector $\boldsymbol{e_{com}} = \sum_{i=1}^N\tilde{m}_i\boldsymbol{e_i}$ is always conserved \citep{Goldreich81}, and therefore always corresponds to one of the Laplace-Lagrange eigenmodes, with zero eigenfrequency.
For an isolated pair of planets, the remaining eigenmode corresponds to the relative eccentricity vector $\boldsymbol{e_{12}} = \boldsymbol{e_{2}}-\boldsymbol{e_{1}}$, which precesses on a secular timescale of $\mathcal{O}(M_\star/m_{tot}) e_{c,12}^2$ orbital periods (Eq.\:\ref{eq:omega}), where $e_{c,12} = \eta a/a_2$ is the eccentricity at which the orbits would cross.

We used this intuition to understand the case where a third massless planet is inserted between a pair of massive planets, since then $\boldsymbol{e_{com}}$ and $\boldsymbol{e_{13}}$ identified above remain unaltered (as far as the two massive planets are concerned, nothing has changed).
In this case we obtained a general expression for the third mode $\boldsymbol{e_-}$, which depends on both how the mass is distributed between the inner and outer planet, and the spacing asymmetry parameter $\eta$ (Eq.\:\ref{eq:eminus}).
For $\eta \rightarrow -1$, corresponding to fixing the inner/outer planet spacing and placing the test particle right next to the inner planet, the close pair form effectively a two-planet subsystem, and $\boldsymbol{e_-} \rightarrow \boldsymbol{e_{12}}$, while for $\eta \rightarrow +1$ when the test particle is next to the outer planet, $\boldsymbol{e_-} \rightarrow \boldsymbol{e_{23}}$.
For equal spacing ($\eta = 0$), $\boldsymbol{e_-} = \tilde{m}_3 \boldsymbol{e_{23}} - \tilde{m}_1 \boldsymbol{e_{12}}$ becomes a simple mass weighted combination of the two relative eccentricities.
In all cases, this $\boldsymbol{e_-}$ mode precesses on approximately the secular timescale given above, with the crossing eccentricity taken relative to the closest neighbor.

This provides a simple interpretation for a geometric picture in which the individual or relative eccentricities are the vector sum of contributions from each of the eigenmodes that each precess at their own rates (Fig.\:\ref{fig:3planets}, Eqs.\:\ref{eq:StoE}-\ref{eq:StoErelative}).
It can also be used to easily estimate the maximum and minimum eccentricities when the modes become aligned or anti-aligned.
The fact that all individual eccentricities carry the same contribution of $\boldsymbol{e_{com}}$ implies that the relative eccentricities $\boldsymbol{e_{ij}}$ between a pair of planets, which are the particular combinations that determine MMR widths \citep{Hadden19}, only have contributions from at most $N-1$ modes.
In particular this implies that the magnitude of $\boldsymbol{e_{12}}$ in two-planet systems is conserved, while in three-planet systems, $\boldsymbol{e_{12}}$ and $\boldsymbol{e_{23}}$ will have contributions from both the $\boldsymbol{e_{-}}$ and $\boldsymbol{e_{13}}$ modes and vary with time.

We then explored how this limiting expression for $\boldsymbol{e_{-}}$ performed as we relaxed our assumptions.
We found that even going out to period ratios of 3:2 between adjacent planets, the error in the mode remained below $15\%$ (Fig.\:\ref{f:varyAlpha13}).
If we instead vary the middle planet's mass from the test-particle to the equal-mass case, the error remains below $10\%$ (Fig.\:\ref{f:varyEta}).
We derive the leading order correction to the mode, which scales as $\tilde{m_2}\eta$ (Eq.\:\ref{eq:correction}).
The fact that both these parameters are typically small, i.e., spacings are roughly uniform so $\eta \ll 1$ and $\tilde{m_2} < 1/3$ (equal masses), and that our test-particle result has the right asymptotic behavior for $\eta$ far from zero (Fig.\:\ref{f:varyEta}), makes our basic result (Eq.\:\ref{eq:eminus}) particularly useful and simple.

We expect that these results will be useful for analytical studies of secular evolution in general, and the secular modulation of MMR boundaries in particular.

\appendix
\section{\label{app:laplag}Laplace-Lagrange Hamiltonian}
Expressed in terms of orbital elements, the Hamiltonian $\mathcal{H}$ for a three-planet system is
\begin{equation}
\mathcal{H} = -\frac{Gm_1m_2}{a_2}\mathcal{R}_{sec,12} - \frac{Gm_2m_3}{a_3}\mathcal{R}_{sec,23} - \frac{Gm_1m_3}{a_3}\mathcal{R}_{sec,13}, \label{eq:Hamiltonian}
\end{equation}
where, to the leading order in eccentricities, the secular disturbing functions between a pair of planets is \citep{Murray99},
\begin{equation}
R_{sec,ij} = \frac{\alpha_{ij}}{8}\left[b^{(1)}_{3/2}(\alpha_{ij})(e_i^2+e_j^2) \right. -\left.2b^{(2)}_{3/2}(\alpha_{ij})e_ie_j\cos(\varpi_j-\varpi_i)\right], \label{eq:R_sec}
\end{equation}
the $\alpha_{ij} \equiv a_i/a_j$ are the semimajor axis ratios, and $b^{(m)}_s$ are Laplace coefficients evaluated at $\alpha_{ij}$.
While in our final expressions we will approximate the $\alpha_{ij} \approx 1$ in the compact limit, we retain them in this Appendix for reference. 
These disturbing functions can be written in matrix notation and in terms of complex eccentricities (Eq.\:\ref{eq:ei}) as

\begin{equation}
R_{sec,ij} = \frac{\alpha_{ij}}{8} \begin{pmatrix} \boldsymbol{e_i^*} & \boldsymbol{e_j^*} \end{pmatrix}
\begin{pmatrix}
b_{3/2}^{(1)}(\alpha_{ij}) & -b_{3/2}^{(2)}(\alpha_{ij}) \\
-b_{3/2}^{(2)}(\alpha_{ij}) & b_{3/2}^{(1)}(\alpha_{ij})
\end{pmatrix}
\begin{pmatrix}\boldsymbol{e_i} \\ \boldsymbol{e_j}\end{pmatrix}. \label{Rsec}
\end{equation}

Our goal is to write the Hamiltonian in terms of the canonical complex $\mathbf{G_i}$ variables.
If we just take the terms for the outermost pair of planets, rewriting the prefactor and taking $\mathbf{e_i^*} = \mathbf{G_i}/\sqrt{\Lambda_i}$, we have
\begin{align}
\mathcal{H}_{23} &=& -\frac{\Lambda_2\Lambda_3\alpha_{23}}{8a_3M_\star\sqrt{a_2 a_3}} \begin{pmatrix} \frac{\boldsymbol{G_2}}{\sqrt{\Lambda_2}} & \frac{\boldsymbol{G_3}}{\sqrt{\Lambda_3}} \end{pmatrix}
\begin{pmatrix}
b_{3/2}^{(1)}(\alpha_{23}) & -b_{3/2}^{(2)}(\alpha_{23}) \\
-b_{3/2}^{(2)}(\alpha_{23}) & b_{3/2}^{(1)}(\alpha_{23})
\end{pmatrix}
\begin{pmatrix}\frac{\boldsymbol{G_2^*}}{\sqrt{\Lambda_2}} \\ \frac{\boldsymbol{G_3^*}}{\sqrt{\Lambda_3}}\end{pmatrix} \nonumber \\
 &=& -\sqrt{\frac{GM_\star}{a_3^3}}\frac{1}{8M_\star} \begin{pmatrix} \boldsymbol{G_1} & \boldsymbol{G_2} & \boldsymbol{G_3} \end{pmatrix}
\begin{pmatrix}
0 & 0 & 0 \\
0 & b_{3/2}^{(1)}(\alpha_{23}) \alpha_{23}^{1/2} m_3 & -b_{3/2}^{(2)}(\alpha_{23}) \alpha_{23}^{3/4} \sqrt{m_2 m_3} \\
0 & -b_{3/2}^{(2)}(\alpha_{23})\alpha_{23}^{3/4} \sqrt{m_2 m_3} & b_{3/2}^{(1)}(\alpha_{23}) \alpha_{23} m_2
\end{pmatrix}
\begin{pmatrix} \boldsymbol{G_1^*} \\ \boldsymbol{G_2^*} \\ \boldsymbol{G_3^*}\end{pmatrix}, \label{eq:Hij}
\end{align}
where in the second equality we additionally expand our matrices to include $\mathbf{G_1}$ (with corresponding zero entries) to make it easier to combine this expression with the corresponding $\mathcal{H}_{ij}$ for the other two pairs of planets.
The contributions $\mathcal{H}_{23}$ and $\mathcal{H}_{13}$ can be easily found by relabeling the inner and outer planets using Eq.\:\ref{eq:Hij} and shifting the entries in the large middle matrix to pick out the appropriate $\mathbf{G_i}$;
however, while $\mathcal{H}_{13}$ shares the same prefactor proportional to the outer body's mean motion $n_3 = \sqrt{GM_\star/a_3^3}$, $\mathcal{H}_{12} \propto n_2 = n_3 \alpha_{23}^{-3/2}$.
If we choose to pull out a common factor $n_3$ from all pairs, and additionally introduce masses scaled by the total planetary mass, $\tilde{m}_i \equiv m_i /m_{tot}$, we finally obtain
\begin{equation}
\mathcal{H} = - \frac{1}{2}\begin{pmatrix}\boldsymbol{G_1} & \boldsymbol{G_2} & \boldsymbol{G}_3\end{pmatrix}
 \cdot{\mathcal{M}}\cdot
\begin{pmatrix}\boldsymbol{G_1^*} \\ \boldsymbol{G_2^*} \\ \boldsymbol{G_3^*}\end{pmatrix},
\end{equation}
where
\begin{equation}
\resizebox{\hsize}{!}{$\mathcal{M} = \frac{n_3}{4}\frac{m_{tot}}{M_\star}
\begin{pmatrix}
\alpha_{12}^{1/2}\alpha_{23}^{-3/2}b^{(1)}_{3/2}(\alpha_{12})\,\tilde{m}_2 + \alpha_{13}^{1/2}b^{(1)}_{3/2}(\alpha_{13})\,\tilde{m}_3 & -\alpha_{12}^{3/4}\alpha_{23}^{-3/2}b^{(2)}_{3/2}(\alpha_{12})\,\sqrt{\tilde{m}_1\tilde{m}_2} & -\alpha_{13}^{3/4} b^{(2)}_{3/2}(\alpha_{13})\,\sqrt{\tilde{m}_1\tilde{m}_3}\\
-\alpha_{12}^{3/4}\alpha_{23}^{-3/2}b^{(2)}_{3/2}(\alpha_{12})\,\sqrt{\tilde{m}_1\tilde{m}_2} & \alpha_{12}\alpha_{23}^{-3/2}b^{(1)}_{3/2}(\alpha_{12})\,\tilde{m}_1+\alpha_{23}^{1/2} b^{(1)}_{3/2}(\alpha_{23})\,\tilde{m}_3 & -\alpha_{23}^{3/4}b^{(2)}_{3/2}(\alpha_{23})\,\sqrt{\tilde{m}_2\tilde{m}_3}\\
-\alpha_{13}^{3/4}b^{(2)}_{3/2}(\alpha_{13})\,\sqrt{\tilde{m}_1\tilde{m}_3} & -\alpha_{23}^{3/4}b^{(2)}_{3/2}(\alpha_{23})\,\sqrt{\tilde{m}_2\tilde{m}_3} & \alpha_{13} b^{(1)}_{3/2}(\alpha_{13})\,\tilde{m}_1 + \alpha_{23} b^{(1)}_{3/2}(\alpha_{23})\,\tilde{m}_2
\end{pmatrix},$} \label{eq:M}
\end{equation}

\section{\label{app:Mprime}Approximated Eigenvalues}
The rotated Laplace-Lagrange matrix given in Eq.\:\ref{eq:M'} is
\begin{equation}
    \mathcal{M'} = \begin{pmatrix}\omega'_1 & k & 0\\
    k & \omega'_2 & 0\\
    0 & 0 & 0 \end{pmatrix} + \delta\omega'_1\mathcal{M}_d, 
\end{equation}
where
\begin{equation}
\begin{aligned}
    \omega'_1 = \tfrac{n_3}{4}\tfrac{m_{tot}}{M_\star}&\left\{\tfrac{\tilde{m}_1}{\tilde{m}_1+\tilde{m}_3} \alpha_{12}^{1/2}\alpha_{23}^{-3/2}\left[\tilde{m}_2^2 b^{(1)}_{3/2}(\alpha_{12}) + (\tilde{m}_1+\tilde{m}_3)^2 \alpha_{12}^{1/2}b^{(1)}_{3/2}(\alpha_{12}) +2\tilde{m}_2(\tilde{m}_1+\tilde{m}_3) \alpha_{12}^{1/4}b^{(2)}_{3/2}(\alpha_{12})\right] \right.\\
    & \left. +\tfrac{\tilde{m}_3}{\tilde{m}_1+\tilde{m}_3}\alpha_{23}\left[\tilde{m}_2^2 b^{(1)}_{3/2}(\alpha_{23}) + (\tilde{m}_1+\tilde{m}_3)^2 \alpha_{23}^{-1/2}b^{(1)}_{3/2}(\alpha_{23}) +2\tilde{m}_2(\tilde{m}_1+\tilde{m}_3) \alpha_{23}^{-1/4}b^{(2)}_{3/2}(\alpha_{23})\right]\right\},
\end{aligned}
\end{equation}

\begin{equation}
\begin{aligned}
    \omega'_2 = \tfrac{n_3}{4}\tfrac{m_{tot}}{M_\star}
    &\left\{\tfrac{\tilde{m}_2}{\tilde{m}_1+\tilde{m}_3} 
    \left(\tilde{m}_3 \alpha_{12}^{1/2}\alpha_{23}^{-3/2}b^{(1)}_{3/2}(\alpha_{12})
    + \tilde{m}_1 \alpha_{23}b^{(1)}_{3/2}(\alpha_{23})\right)\right.\\
    &\left.+ \tfrac{1}{\tilde{m}_1+\tilde{m}_3}
    \alpha_{13}\left(\tilde{m}_1^2b^{(1)}_{3/2}(\alpha_{13})+\tilde{m}_3^2\alpha_{13}^{-1/2}b^{(1)}_{3/2}(\alpha_{13}) + 2\tilde{m}_1\tilde{m}_3\alpha_{13}^{-1/4}b^{(2)}_{3/2}(\alpha_{13})\right)\right\},
\end{aligned}
\end{equation}

\begin{equation}
\begin{aligned}
    k = \tfrac{n_3}{4}\tfrac{m_{tot}}{M_\star}\tfrac{\sqrt{\tilde{m}_1\tilde{m}_2\tilde{m}_3}}{\tilde{m}_1+\tilde{m}_3}
    &\left\{\alpha_{23}\left[\tilde{m}_2 b^{(1)}_{3/2}(\alpha_{23}) + (\tilde{m}_1+\tilde{m}_3)\alpha_{23}^{-1/4}b^{(2)}_{3/2}(\alpha_{23})\right]\right.\\
    &\left.-\alpha_{12}^{1/2}\alpha_{23}^{-3/2}\left[\tilde{m}_2 b^{(1)}_{3/2}(\alpha_{12}) + (\tilde{m}_1+\tilde{m}_3)\alpha_{12}^{1/4}b^{(2)}_{3/2}(\alpha_{12})\right]\right\}.
\end{aligned}
\end{equation}

If we assume that $b^{(1)}_{3/2}(\alpha_{ij}) = b^{(2)}_{3/2}(\alpha_{ij})$ and $\alpha_{ij}^{1/4}=1$ in the square brackets in order to combine terms, the expressions reduce to
\begin{equation}
    \omega'_1 = \frac{n_3}{2\pi}\frac{m_{tot}}{M_\star}
    \left(\frac{\tilde{m}_1}{\tilde{m}_1+\tilde{m}_3}\frac{1}{e_{c,12}^2}
    + \frac{\tilde{m}_3}{\tilde{m}_1+\tilde{m}_3}\frac{1}{e_{c,23}^2}\right), \label{eq:w1}
\end{equation}
\begin{equation}
    \omega'_2 = \frac{n_3}{2\pi}\frac{m_{tot}}{M_\star}\left[
    \tilde{m}_2\left(\frac{\tilde{m}_3}{\tilde{m}_1+\tilde{m}_3}\frac{1}{e_{c,12}^2} + \frac{\tilde{m}_1}{\tilde{m}_1+\tilde{m}_3}\frac{1}{e_{c,23}^2}\right) + (\tilde{m}_1+\tilde{m}_3)\frac{1}{e_{c,13}^2}\right], \label{eq:w2}
\end{equation}
\begin{equation}
    k = \frac{n_3}{2\pi}\frac{m_{tot}}{M_\star}\frac{\sqrt{\tilde{m}_1\tilde{m}_2\tilde{m}_3}}{\tilde{m}_1+\tilde{m}_3}\left(\frac{1}{e_{c,23}^2} - \frac{1}{e_{c,12}^2}\right), \label{eq:k}
\end{equation}
where
\begin{equation}
    e_{c,12} \equiv \alpha_{12}^{-1/4}\alpha_{23}^{5/8}(1-\alpha_{12}), \quad
    e_{c,23} \equiv \alpha_{12}^{1/8}\alpha_{23}^{-1/2}(1-\alpha_{23}), \quad
    e_{c,13} \equiv \alpha_{13}^{-1/2}(1-\alpha_{13}). \label{eq:ecrosses}
\end{equation}

While we retain these more accurate expressions for reference, for simplicity in the main text we set the $\alpha_{ij}$ prefactors in Eq.\:\ref{eq:ecrosses} to unity in generating our plots as appropriate in the compact limit. 

\bibliography{Bib}

\bibliographystyle{aasjournal}
\end{document}